\documentclass{svproc}
\usepackage[a4paper,margin=2cm]{geometry}
\usepackage{graphicx, multirow, outlines, ulem}
\usepackage[utf8]{inputenc}
\usepackage{amsmath} 
\usepackage{amsfonts}
\usepackage{amssymb}
\usepackage{caption}
\usepackage{subcaption}
\usepackage{float}
\usepackage{cite}
\usepackage{epstopdf}
\DeclareGraphicsExtensions{.pdf,.png,.jpg}
\usepackage{breqn}
\usepackage{upgreek}

\usepackage{xcolor}
\colorlet{linkequation}{red}
\usepackage[colorlinks]{hyperref}

\begin{document}
\mainmatter              
\title{Modified Hellmann Feynman Theorem}
\titlerunning{MHFT}  
%
\author{Gaurav Hajong\inst{1} \and
Bhabani Prasad Mandal\inst{1}}
\authorrunning{Gaurav Hajong et al.} 
%
\tocauthor{Gaurav Hajong and BP Mandal}
\institute{BHU, Varanasi 221005, India\\
\email{gauravhajong730@gmail.com}
}
\maketitle              

\begin{abstract}
 We review the well-known Hellmann Feynman Theorem (HFT), originally developed for Hermitian systems to facilitate the calculation of forces among the molecules. Our work extends this foundational theorem to the domain of non-Hermitian quantum mechanics, in particular the $PT$ symmetric non-Hermitian quantum physics. We derive a modified form of the HFT (MHFT) which holds good for both $PT$ broken, unbroken phases and even at the exceptional point of the theory as demonstrated with help of a discrete and a continumm model. Since a $PT$-symmetric Hamiltonian admits biorthonormal set of eigenvectors, a more appropriate inner product known as the $G$ inner product is defined, based on which, the system in the unbroken phase can be shown to satisfy unitary time evolution, while a system in broken phase does not. We show here that the MHFT obtained is valid for both these situations.
\end{abstract}
\section{Introduction}
The well-known Hellmann Feynman Theorem  (HFT) \cite{PhysRev.56.340}, which was formulated as a means to calculate the forces in molecules in equilibrium is stated as follows 
\begin{equation}\label{hft}\frac{\partial E_\lambda}{\partial\lambda}=\Big\langle\psi_\lambda \Big | \frac{\partial \widehat{H}_\lambda}{\partial\lambda}\Big | \psi_\lambda\Big\rangle,
\end{equation}
where $\lambda $ is some arbitrary parameter in the Hamiltonian and $| \psi_\lambda\rangle $ is an eigenstate of the system. The theorem has found its place in various areas of physics ever since, including high energy physics \cite{FERNANDEZ2022169158,PhysRevD.101.094508}, condensed matter physics \cite{PhysRevB.104.L100506,De_Rosi_2023} and machine learning techniques \cite{doi:10.1021/acs.chemrev.0c01111}. 

 For quite some time, a special class of quantum systems, popularly known as the $PT$-symmetric non-Hermitian system, have found its place secured in almost all branches of frontier research of physics. The system is non-Hermitian in the sense that the usual self-adjoint condition on the Hamiltonian is replaced by a much more  physical and rather less constraining condition of $PT$-symmetry~\cite{Bender_2002,KHARE200053}. The importance of such non-Hermitian system lies in the fact that it possesses real eigenvalue spectrum over a certain range of the non-hermitian parameter. Basically, it can be divided into two categories: Unbroken phase and Broken phase. Unbroken phase is the one in which the  eigenvectors are $PT$-symmetric and the entire eigenvalue spectrum is real and the broken phase is the one in which the system has at least one complex conjugate pair of eigenvalues and the eigenvector(s) are not $PT$-symmetric \cite{RAVAL2019114699,MANDAL2015185}. The transition point between these two phases turns out to be the exceptional point (EP) of the system where the eigenvalues as well as the eigenvectors coalesce. While for the unbroken phase, a consistent quantum theory having real eigenvalue spectrum, unitary time evolution and positive norm can be established in a modified Hilbert space equipped with an appropriate positive definite inner product~\cite{PhysRevA.100.062118,un_23}, the broken phase is not clearly understood as for this phase the unitary evolution is not satisfied.
This opens up an opportunity for further studies in the broken region.

The next section is aimed at finding the suitable metric operator based on which consistent quantum theory can be established.
\section{Biorthonormal System}

We first consider a general $PT$-invariant non-Hermitian Hamiltonian, $\widehat{\mathcal{H}}$ defined as, $\widehat{\mathcal{H}}^\dag\not=\widehat{\mathcal{H}},\ \  \big[PT,\widehat{\mathcal{H}}\big]=0$. Such systems are characterized by right eigenvectors $|R_i\rangle $ and left eigenvectors 
$|L_i\rangle$, defined as
$$\widehat{\mathcal{H}}|R_i\rangle=E_i|R_i\rangle, \ \ \  \widehat{\mathcal{H}}^\dag |L_i\rangle=E^*_i|L_i\rangle.  $$
where, $|L_i\rangle$ and $|R_i\rangle$ are related to each other by a positive definite hermitian metric operator $G$ as $\langle{L_i}|=\langle{R_i|G}$. The metric operator $G$ is defined as \cite{un_23} \begin{equation} \label{G_operator} G= \sum_i |L_i\rangle\langle L_i| = \left [ \sum_i |R_i\rangle\langle R_i|\right ]^{-1}.
\end{equation}
Also, $|L_i\rangle$ and $|R_i\rangle$ form a complete  bi-orthogonal set \cite{kleefeld2009construction}. Thus, the completeness relation and the general inner product \cite{PhysRevA.100.062118} are formulated as Eq.(\ref{complete}) and (\ref{ortho_com_rel}) respectively.  
\begin{equation} \label{complete}\sum_i|R_i\rangle\langle{L_i}| =I
 \end{equation}
 \begin{equation} \label{ortho_com_rel}\langle{L_i}|R_j\rangle=\langle{R_i}|G|R_j\rangle = \langle{R_i}|R_j\rangle_G = \delta_{ij}.
 \end{equation} 
The definition of expectation value of an observable $O$ will also be redefined with respect to the G-inner product as, 
\begin{equation}
 \label{G_expt}\langle O\rangle_G=\langle{R_i}|GO|R_i\rangle= \langle{L_i}|O|R_i\rangle.
 \end{equation}
It can be shown that $\langle O\rangle_G$ is a real number \cite{un_23,PhysRevA.100.062118,mostafazadeh2010pseudo} for any state in the Hilbert space if and only if $O$ satisfies the following condition i.e. 
$$O^\dag G = G O.$$
The observables which obey the above condition are called ``good observables".

Supported by a rigorous derivation, it has been shown that for a generic non-Hermitian system, the Hellmann Feynman Theorem takes the form \cite{PhysRevA.109.022227}
\begin{eqnarray}\label{hft_unbroken_1} 
 \frac{\partial E_\lambda}{\partial\lambda}=  \Big\langle{L}\Big|\frac{\partial\widehat{\mathcal{H}}}{\partial\lambda}\Big|R\Big\rangle=\Big\langle{R}\Big|G\frac{\partial\widehat{\mathcal{H}}}{\partial\lambda}\Big|R\Big\rangle.
 \end{eqnarray}
 which we call as the modified Hellmann-Feynman Theorem (MHFT).
 
 We will now consider explicit examples to further confirm the validity of this theorem.
\section{Discrete model}

We consider a $PT$-symmetric non-Hermitian two-level system as proposed by Wang \cite{doi:10.1098/rsta.2012.0045}, described by the following Hamiltonian, 
 \[
H_{2\times2} = \left( \begin{matrix} \epsilon+\gamma\cos\delta & -i\left(\gamma\sin\delta-\rho\right) \\ i\left(\gamma\sin\delta+\rho\right) & \epsilon-\gamma\cos\delta \end{matrix} \right)
\]
This Hamiltonian is $PT$-symmetric under the action of parity operator,

\[
P = \left( \begin{matrix} 1 & 0 \\ 0 & -1 \end{matrix} \right)
\]
The eigenvalues of $H_{2\times{2}}$ are $E_\pm=\epsilon\pm\sqrt{\gamma^{2}-\rho^2}$. Clearly, this system undergoes a $PT$ phase transition at $\rho = \gamma$. Firstly, we consider only the unbroken phase. With a suitable parametrization we try to find the eigenvectors.

Let $\frac{\rho}{\gamma}=\sin\alpha$, such that $\rho\leq\gamma$, therefore, the eigenvalues can be rewritten as, \begin{equation}\label{eigenvl}E_\pm=\epsilon\pm\gamma\cos\alpha
\end{equation}
and the corresponding right eigenvectors can be determined to be,
$$|R_{+}\rangle=\frac{1}{\sqrt{\cos\alpha}}\begin{pmatrix}{\cos\left(\frac{\delta+\alpha}{2}\right)}\\ {i\sin\left(\frac{\delta+\alpha}{2}\right)}
\end{pmatrix},$$ $$|R_{-}\rangle=\frac{1}{\sqrt{\cos\alpha}}\begin{pmatrix}{i\sin\left(\frac{\delta-\alpha}{2}\right)}\\ {\cos\left(\frac{\delta-\alpha}{2}\right)}
\end{pmatrix},$$
The left eigenvectors are determined as, 
$$|L_{+}\rangle=\frac{1}{\sqrt{\cos\alpha}}\begin{pmatrix}{\cos\left(\frac{\delta-\alpha}{2}\right)}\\ {i\sin\left(\frac{\delta-\alpha}{2}\right)}
\end{pmatrix},$$ $$|L_{-}\rangle=\frac{1}{\sqrt{\cos\alpha}}\begin{pmatrix}{i\sin\left(\frac{\delta+\alpha}{2}\right)}\\ {\cos\left(\frac{\delta+\alpha}{2}\right)}
\end{pmatrix}.$$ 
Now it can be shown for $|R_+\rangle$ and $|R_-\rangle$ in the unbroken phase that
\begin{equation}
\label{+-revur} 
\Big\langle{L_{\pm}}\Big|\frac{\partial{H_{2\times2}}}{\partial\lambda}\Big|R_{\pm}\Bigr\rangle=\mp\frac{\rho}{\sqrt{\gamma^2-\rho^2}} = \frac{\partial{E_{\pm}}}{\partial\lambda}.
\end{equation}
Similarly, for broken phase $(\rho>\gamma)$, one obtains, 
\begin{equation}
\label{+revur1} 
\Big\langle{L^b_{\pm}}\Big|\frac{\partial{H_{2\times2}}}{\partial\rho}\Big|R^b_{\pm}\Bigr\rangle={\pm}i\frac{\rho}{\sqrt{\rho^2-\gamma^2}} = \frac{\partial{E^*_{\pm}}}{\partial\rho},
\end{equation}
where $|L^b\rangle$ are $|R^b\rangle$ are the eigenvectors in the broken phase.

It is important to note that a PT-symmetric non-hermitian system is non-unitary in the broken region, i.e., $$\langle{R_i}(t)|G_b|R_i(t)\rangle \neq\langle{R_i}|G_b|R_i\rangle.$$ 
However, we can instead show that $$\langle{R_i}(t)|G_b(t)|R_i(t)\rangle =\langle{R_i}|G_b|R_i\rangle = 1,$$ where $G_b(t)$ \cite{PhysRevResearch.3.013015} is time-dependent. This requires us to check for the validity of Eqn.(\ref{hft_unbroken_1}) by considering the time-dependent states as well.

As for the discrete system in \cite{PhysRevA.103.062416}, MHFT has been shown to be working well in the broken region for the eigenstates at time, $t=0$ with time independent $G_b$ \cite{PhysRevA.109.022227}, here, we only show the validity of the theorem in the broken region using time dependent $G_b(t)$ for the eigenstates under time evolution .
It turns out that under time evolution too, we obtain the following relation to be consistent, i.e.,\begin{equation}
\label{+te} 
\Big\langle{L^b_{\pm}(t)}\Big|\frac{\partial{H_{2\times2}}}{\partial\rho}\Big|R^b_{\pm}(t)\Big\rangle=\Big\langle{R^b_{\pm}(t)}\Big|G_{b}(t)\frac{\partial{H_{2\times2}}}{\partial\rho}\Big|R^b_{\pm}(t)\Big\rangle={\pm}i\frac{\rho}{\sqrt{\rho^2-\gamma^2}} = \frac{\partial{E^*_{\pm}}}{\partial\rho},
\end{equation}
This confirms the validity of the MHFT for this model.

\section{Continuum Model}
\begin{figure}[h!]   
    \centering\includegraphics[width=0.48\textwidth]{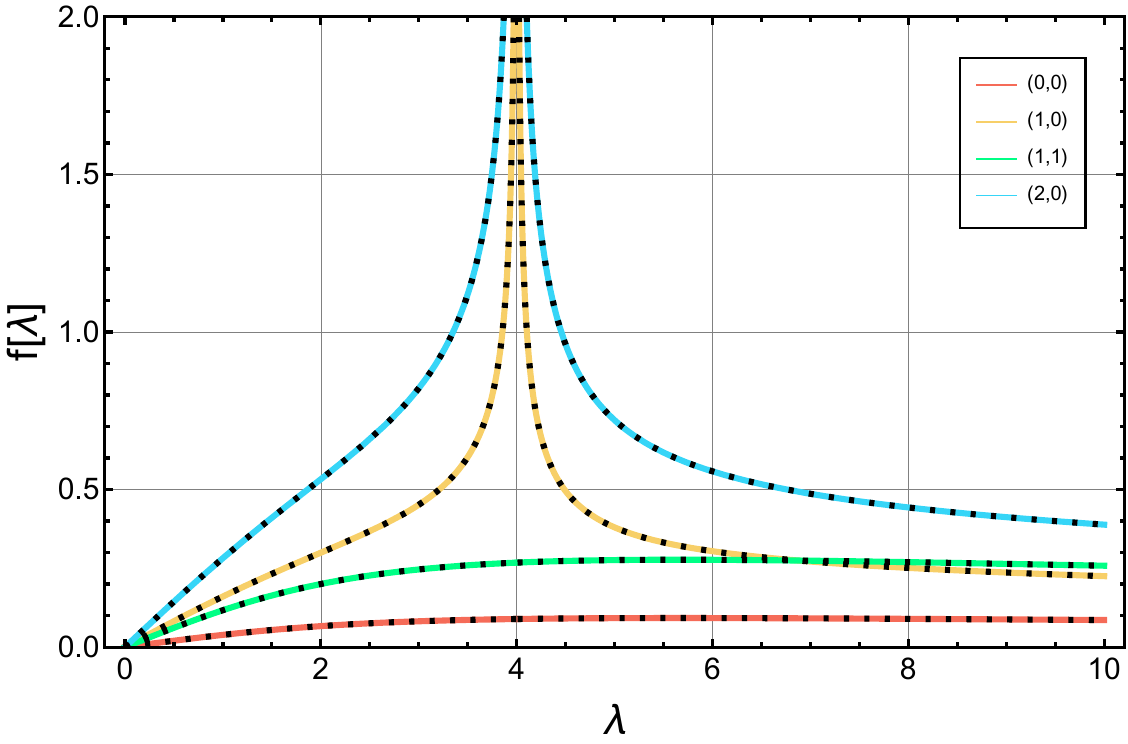}
    \caption{Comparison between the absolute value of $\Big\langle\frac{\partial\widehat{\mathcal{H}}}{\partial\lambda}\Big\rangle_G$ and $\frac{\partial{E_{n_1,n_2}}}{\partial\lambda}$ for both broken and unbroken regions for the states $(n_1,n_2) = (0,0), (1,0), (1,1)$ and $(2,0)$. The  solid lines and black dotted lines indicate the LHS  and RHS of Eq.~\eqref{hft_integral}.  }
    \label{fig:1,0*abo}
\end{figure}

As it can be seen from \cite{PhysRevA.109.022227}, that for the system of a 2-d anharmonic oscillator with a  non-Hermitian interaction term \cite{MANDAL20131043}, described by 

\begin{equation}
\label{pt_hamiltonian}
   H_{2d}=\frac{p_x^2}{2m}+\frac{p_y^2}{2m}+\frac{1}{2}m\omega_x^2x^2+\frac{1}{2}m\omega_y^2y^2+i\lambda{xy},
\end{equation}
where $\lambda$ is real and $\omega_x\neq\omega_y$, the energy eigenvalues and the right and left eigenvectors can be obtained as, 

\begin{align}
E_{n_1,n_2} &= \left(n_1+\frac{1}{2}\right)\hbar{C_1} + \left(n_2+\frac{1}{2}\right)\hbar{C_2}, \\
R_{n_1,n_2} &= N e^{-\frac{m}{2\hbar}\left[C_1 X^2 + C_2 Y^2\right]} H_{n_1}\left(\alpha_1 X\right) H_{n_2}\left(\alpha_2 Y\right), \\
L_{n_1,n_2} &= N e^{-\frac{m}{2\hbar}\left[C_1^* {X^*}^2 + C_2^* {Y^*}^2\right]} H_{n_1}\left(\alpha_1^* X^*\right) H_{n_2}\left(\alpha_2^* Y^*\right).
\end{align}
In integral form, the MHFT in the case of continuum models has been shown to satisfy,  

\begin{eqnarray}
\label{hft_integral}
\frac{\int{{\Big(L_{n_1,n_2}^*}\Big)\frac{\partial{H_{2d}}}{\partial\lambda}\Big(R_{n_1,n_2}\Big)dxdy}}{\int{{\Big(L_{n_1,n_2}^*}\Big)\Big(R_{n_1,n_2}\Big)dxdy}}=\frac{\partial{E_{n_1,n_2}}}{\partial\lambda}
\end{eqnarray}
For $n_1= n_2$, we have real eigenvalues, with the eigenvectors being PT-symmetric over all values of $\lambda$.

Fig.(\ref{fig:1,0*abo}) represents the absolute values of the results produced from Eq.(\ref{hft_integral}) for the ground state $(0,0)$ and various other low lying excited states, covering both the unbroken and broken regions ($\lambda = 4$ being the exceptional point). The solid and the dotted lines correspond to the L.H.S. and R.H.S. of Eq.(\ref{hft_integral}). Explicit verification of the theorem for the real and the imaginary parts have been done in \cite{PhysRevA.109.022227}. 

\section{Results and Conclusions}
We find in this work that our derivation of the MHFT for the non-Hermitian systems indeed works for PT-symmetric discrete models in both the PT unbroken and the broken phases. We have also verified its validity for a continuum model as well. An example for a non-PT symmetric non-Hermitian system has been provided in \cite{PhysRevA.109.022227}. Having established this form of MHFT, we can look for calculating force in the context of non-Hermitian systems \cite{PhysRevE.111.014421} like in the case of open quantum systems, where a suitable choice of the parameter has to be made.

\section{Acknowledgement}

BPM acknowledges the incentive research grant for faculty under the IoE Scheme (IoE/Incentive/2021-22/32253 of Banaras Hindu University. GH acknowledeges UGC, New Delhi, India for JRF Fellowship.)
%
%


\begin{thebibliography}{6}
%
\bibitem {PhysRev.56.340}{R. P. Feynman}, {Phys. Rev.}, {1939}, {10.1103/PhysRev.56.340}

\bibitem {PhysRevB.104.L100506}{Joseph D. Pakizer, and Alex Matos-Abiague},  {Phys. Rev. B}, {104}, {2021}, {10.1103/PhysRevB.104.L100506}

\bibitem {De_Rosi_2023}
{Giulia De Rosi and Riccardo Rota and Grigori E Astrakharchik    and Jordi Boronat}, {10.1088/1367-2630/acc6e6}, 2023, {{IOP} Publishing}

\bibitem {FERNANDEZ2022169158}
Francisco M. Fernández. Annals of Physics, 447,{2022},
{https://doi.org/10.1016/j.aop.2022.169158}

\bibitem {PhysRevD.101.094508}
{Ra\'ul A. Brice\~no, Maxwell T. Hansen, and Andrew W. Jackura},
{Phys. Rev. D}, {2020}, {10.1103/PhysRevD.101.094508}

\bibitem {doi:10.1021/acs.chemrev.0c01111} {Oliver T. Unke, Stefan Chmiela, Huziel E. Sauceda, and Michael Gastegger, Igor Poltavsky, Kristof T. Schütt, Alexandre Tkatchenko and Klaus-Robert Müller}, {Chemical Reviews}, {121}, {2021}, {10.1021/acs.chemrev.0c01111}


\bibitem {Bender_2002} {Carl M. Bender, Dorje C. Brody, and Hugh F. Jones},  {Phys. Rev. Lett.}, {89},{2002},{10.1103/PhysRevLett.89.270401}

\bibitem {KHARE200053} {Avinash Khare and Bhabani Prasad Mandal}, {Physics Letters A}, {272}, {2000}, {https://doi.org/10.1016/S0375-9601(00)00409-6}

\bibitem {MANDAL2015185}{B.P. Mandal and B.K. Mourya and K. Ali and A. Ghatak}, {Annals of Physics}, {363}, {2015}, {https://doi.org/10.1016/j.aop.2015.09.022}

\bibitem {RAVAL2019114699} {Nuclear Physics B}, {Haresh Raval and Bhabani Prasad Mandal}, {946}, {2019}, {https://doi.org/10.1016/j.nuclphysb.2019.114699}

\bibitem {un_23} {Namrata Shukla, Ranjan Modak and Bhabani Prasad Mandal}, {Phys. Rev. A}, {107}, {042201}, {2023},{10.1103/PhysRevA.107.042201}

\bibitem {PhysRevA.100.062118} {Chia-Yi Ju, Adam Miranowicz, Guang-Yin Chen, Franco Nori}, {Phys. Rev. A}, {100}, {062118}, {2019}, {10.1103/PhysRevA.100.062118}

\bibitem {kleefeld2009construction} {F. Kleefeld}, {}, {2009}, {0906.1011}, {arXiv}, {hep-th}

\bibitem {mostafazadeh2010pseudo} { Ali Mostafazadeh}, {International Journal of Geometric Methods in Modern Physics}, {7}, {2010},{10.1142/S0219887810004816}
 
\bibitem {doi:10.1098/rsta.2012.0045} {Wang, Qing-hai }, {Philosophical Transactions of the Royal Society A: Mathematical, Physical and Engineering Sciences}, {371}, {2013}, {10.1098/rsta.2012.0045}

\bibitem {PhysRevA.109.022227}{Gaurav Hajong, Ranjan Modak and Bhabani Prasad Mandal}, {Phys. Rev. A}, {109}, {2024}, {10.1103/PhysRevA.109.022227}

\bibitem {PhysRevResearch.3.013015}
{Yu-Chin Tzeng, Chia-Yi Ju, Guang-Yin Chen and Wen-Min Huang}, {Phys. Rev. Res.}, {3}, {2021}, {10.1103/PhysRevResearch.3.013015}

\bibitem {MANDAL20131043} {Bhabani Prasad Mandal and Brijesh Kumar Mourya and Rajesh Kumar Yadav}, {Physics Letters A}, {377}, {2013}, {https://doi.org/10.1016/j.physleta.2013.02.023} 

\bibitem {PhysRevA.103.062416} {Ranjan Modak, and Bhabani Prasad Mandal}, {Phys. Rev. A}, {103}, {2021}, {10.1103/PhysRevA.103.062416}

\bibitem {PhysRevE.111.014421} {Tanmoy Pal, Ranjan Modak and Bhabani Prasad Mandal}, {Phys. Rev. E}, {2025}, {10.1103/PhysRevE.111.014421}


\end{thebibliography}
\end{document}